\documentclass[final,3p,times,twocolumn]{elsarticle}

\usepackage[font=footnotesize]{subcaption}
\usepackage{float}
\usepackage{color}
\usepackage{graphics}
\usepackage{graphicx}
\usepackage{amsmath}
\usepackage{amssymb}
\usepackage{microtype}
\usepackage{comment}
\usepackage{multirow}
\usepackage{booktabs}
\usepackage{array}

\usepackage{tikz}
\usepackage{url}
\usepackage{balance}
\usepackage{flushend}
\usepackage[noend]{algpseudocode}
\usepackage{algorithm}
\usepackage{algorithmicx}
\usepackage{stmaryrd}
\usepackage{lineno}

\usepackage{array}
\newcolumntype{M}[1]{>{\centering\arraybackslash}m{#1}}
\newcolumntype{N}{@{}m{0pt}@{}}

\mathchardef\mhyphen="2D
\newcommand{\mse}{\ensuremath{\mathrm{MSE}}}
\newcommand{\psnr}{\ensuremath{\mathrm{PSNR}}}

\journal{}

\let\linenumbers\nolinenumbers\nolinenumbers

\begin{document}

\sloppy
\tolerance = 999

\nolinenumbers

\begin{frontmatter}

\title{Single Image Super-Resolution \\ Based on Capsule Neural Networks}

\author{George Corr\^ea de Ara\'ujo}
\author{Helio Pedrini}
\address{Institute of Computing, University of Campinas, Campinas, SP, 13083-852, Brazil}
\ead{helio@unicamp.br}

\begin{abstract}
Single image super-resolution (SISR) is the process of obtaining one high-resolution version of a low-resolution image by increasing the number of pixels per unit area. This method has been actively investigated by the research community, due to the wide variety of real-world problems where it can be applied, from aerial and satellite imaging to compressed image and video enhancement. Despite the improvements achieved by deep learning in the field, the vast majority of the used networks are based on traditional convolutions, with the solutions focusing on going deeper and/or wider, and innovations coming from jointly employing successful concepts from other fields. In this work, we decided to step up from the traditional convolutions and adopt the concept of capsules. Since their overwhelming results both in image classification and segmentation problems, we question how suitable they are for SISR. We also verify that different solutions share most of their configurations, and argue that this trend leads to fewer explorations of network varieties. During our experiments, we check various strategies to improve results, ranging from new and different loss functions to changes in the capsule layers. Our network achieved good results with fewer convolutional-based layers, showing that capsules might be a concept worth applying in the image super-resolution problem.
\end{abstract}


\end{frontmatter}

\linenumbers





\section{Introduction}
\label{sec:introduction}

In recent years, various techniques have been developed that have advanced the state-of-the-art in image super-resolution. Many of these methods are based on artificial intelligence concepts, more specifically on deep neural networks, and have achieved visually striking results~\cite{zhang2018rcan,yu2018wide,Ren_2020_CVPR_Workshops,Ji_2020_CVPR_Workshops,Behjati_2021_WACV}. However, the vast majority of recent models are composed of traditional convolutional layers which, although widely studied and optimized, have limitations that prevent a better understanding of the data by the network, not absorbing valuable information such as interrelationship between its components.

Recently, Sabour et al.~\cite{NIPS2017_6975} presented an implementation of the capsule concept, a group of neurons whose activation vector represents the parameters that describe a specific entity, such as an object or part of it, and its presence. Using a three-layer capsule network, Sabour et al. achieved results comparable to those of deeper convolutional neural networks for the digit identification problem. The authors also obtained a 5\% error rate on the segmentation of two different digits with an 80\% overlap, a rate previously achieved only in much simpler cases (with less than 4\% overlap between images).

Other authors have also demonstrated the abstraction power of the capsules. Hinton et al.~\cite{46653} reduced by 45\% the error rate on the problem of identifying objects from different perspectives while making the network more resistant to adversarial attacks. LaLonde and Bagci~\cite{lalonde2018capsules} have created a network that can process images tens of times larger while reducing the number of required parameters by 38.4\% and increasing the quality of medical image segmentation.

A few works proposed the usage of capsule-based networks to solve problems that involve low resolution images. Singh et al.~\cite{Singh_2019_ICCV} proposed a Dual Directed Capsule Network model, termed as DirectCapsNet, which utilizes a combination of capsule and convolutional layers for addressing the very low resolution (VLR) digit and face recognition problem. Majdabadi and Ko~\cite{majdabadi2020capsule} implemented a Generative Adversarial Network (GAN) that uses a CapsNet as the discriminator for facial image super resolution, surpassing compared baselines in all metrics. Hsu et al.~\cite{8950449} developed two frameworks, namely Capsule Image Restoration Neural Network (CIRNN) and Capsule Attention and Reconstruction Neural Network (CARNN), to incorporate capsules into image SR convolutional neural networks, achieving better performance than traditional CNN methods with a similar amount of parameters. Despite these capsule usages having demonstrated its usefulness, most of these works are based on the plain CapsNet implementation by Sabour et al.~\cite{NIPS2017_6975}, failing in exploring novel capsule concepts.

This work focuses on the problem of increasing the resolution of varied images, based on a single reference image, through the application of deep learning techniques based on capsules. We implement a neural network based on newer concepts of capsules for single image super-resolution (SISR) problems. In the evaluation of this work, we used datasets publicly available and commonly used as benchmarks, such as Set5~\cite{bevilacqua2012low}, Set14~\cite{5466111}, B100~\cite{937655}, Urban100~\cite{Huang_2015_CVPR}, and the validation set of DIV2K~\cite{Agustsson_2017_CVPR_Workshops}.

\section{Background}
\label{sec:background}

In this section, some relevant concepts related to the topic addressed in our work are described.

\subsection{Super-Resolution}

Super-resolution (SR) is the process of obtaining one or more plausible high resolution images (HR) from one or more low resolution images (LR)~\cite{Nasrollahi2014}. It is an area that has been studied for decades~\cite{Irani:1991:IRI:108693.108696} and has a wide variety of application fields such as aerial imaging, medical image processing, video resolution improvement, traffic sign reading, among others~\cite{Nasrollahi2014,Shi_2016_CVPR}. The relationship between LR and HR images may vary depending on the situation. Many studies assume that the LR image is a reduced version of the HR image by bicubic interpolation, but other degradation factors can be considered in real examples, such as quantization errors, acquisition sensor limitations, presence of noise, blurring, and even the use of different interpolation operators aiming resolution reduction for storage~\cite{8014883}.

The first successful usage of neural networks for SISR problems was developed by Dong et al.~\cite{10.1007/978-3-319-10593-2_13}. With their Super-Resolution Convolutional Neural Network (SRCNN) model, they created a complete solution that maps LR images to SR versions with little pre/post-processing that produced the far best result. After their achievement, several other works have advanced the state-of-the-art in the SISR problem~\cite{Kim_2016_CVPR_2,Kim_2016_CVPR,7115171,7410407}. However, they have strong limitations.

These models receive as input an enlarged version of the LR image, usually through bicubic interpolation, and seek to improve the quality of the image. This means that the operations performed by the neural network are all done in high resolution space, which is inefficient and incurs high processing cost. The computational complexity of the convolution grows quadratically with the size of the input image, whose generation of an SR image with a scaling factor $n$ would result in a cost $n^{2}$ compared to the processing in the low resolution space~\cite{10.1007/978-3-319-46475-6_25}.

Looking for a way of postponing the resolution increase in the network, Shi et al.~\cite{Shi_2016_CVPR} developed a new layer, called the subpixel convolution (or PixelShuffle), which works equivalent to a deconvolution with kernel size divisible by spacing, but is $\log_{2}r^{2}$ times faster compared to deconvolution. Its network, named Efficient Sub-pixel Convolutional Neural Network (ESPCN), achieved speed improvements of over $10\times$ compared to SRCNN~\cite{10.1007/978-3-319-10593-2_13} while having a greater number of parameters and achieving better results for an upscaling factor of $4\times$.

The concept of subpixel convolution is currently the most employed to perform upscaling in neural networks, and has been used by several solutions that have reached the best results~\cite{Ledig_2017_CVPR,10.1007/978-3-030-11021-5_5,Lim_2017_CVPR_Workshops,DBLP:journals/corr/abs-1802-08797,yu2018wide,zhang2018rcan,Ren_2020_CVPR_Workshops,Behjati_2021_WACV} and participated in several editions of the SISR competition that took place during the New Trends in Image Restoration and Enhancement workshop\cite{8014883,Timofte_2018_CVPR_Workshops,Cai_2019_CVPR_Workshops,Zhang_2020_CVPR_Workshops}.

\subsection{Capsules}

Initially introduced by Hinton et al.~\cite{10.1007/978-3-642-21735-7_6}, the concept of capsule proposes to solve some of the main flaws found in traditional convolutional networks: inability to identify spatial hierarchy between elements and lack of rotation invariance. Hinton et al. conclude that, after several stages of subsampling, these networks lose information that make them possible to identify the spatial relationship between the elements of an image.

The authors argue that, contrary to looking for a point of view invariance of the neurons' activities that use a single output value, neural networks should use local ``capsules'' which learn to recognize a visual entity implicitly under a limited domain of viewing conditions and deformations. These structures would encapsulate the result of these complex calculations into a small, highly informative output vector, which could contain information such as the probability of that entity being present in that limited domain, and a set of instantiation parameters, which would include deformation, pose (position, size, orientation), hue, texture, and illumination condition of the visual entity relative to the version learned by the capsule~\cite{NIPS2017_6975}.

Although idealized by Hinton et al., the first successful implementation of the capsule concept was made by Sabour et al.~\cite{NIPS2017_6975}\footnote{Available at \url{https://github.com/Sarasra/models/tree/master/research/capsules}}. In their work, the authors have created a three-layer capsule network that achieved comparable results with the best results in the MNIST~\cite{726791} digit classification problem, which was previously only achieved by deeper networks. For this, they developed two innovative concepts: dynamic routing and a new activation function that acts on a vector output.

After the successful work of Sabour et al.~\cite{NIPS2017_6975}, many other authors have further developed the concept of capsules. Hinton et al.~\cite{46653} proposed a new type of capsule which has a logistic unit that indicates the probability of the presence of an entity and a pose matrix of $4 \times 4$ representing the pose of that entity. The authors also introduced a new routing algorithm, which allows the outputs of the capsules to be routed to those of the next layer so that the active capsules receive a group of votes from similar poses. Hinton et al. showed that their model surpasses the best result in the smallNORB dataset, reducing the number of errors by more than 40\% while being significantly more resistant to white-box adversarial attacks.

A remarkable work, which made possible the development of the proposed solution, was developed by LaLonde and Bagci~\cite{lalonde2018capsules}\footnote{Available at \url{https://github.com/lalonderodney/SegCaps}}. The authors expanded the use of capsules for the problem of object segmentation and made innovations that allowed, among other gains, to increase the data processing capacity of the capsule network, increasing from inputs of $32 \times 32$ to $512 \times 512$ pixels. Most significantly, they were able to advance the state-of-the-art in the problem of segmentation of lung pathologies from computed tomography, while reducing the number of parameters by approximately 38\%. For such, modifications were made to the capsule routing algorithm, as well as in the reconstruction part, and also the concept of convolutional capsules was created.

Recently, a few authors have employed capsules in their solutions for problems involving LR images~\cite{Singh_2019_ICCV, majdabadi2020capsule, 8950449}. It is worth noting that most of these solutions only made small changes to the first capsule networks introduced by Sabour et al.~\cite{NIPS2017_6975} and Hinton et al.~\cite{46653}. In the works that focuses on SR image generation, Majdabadi and Ko~\cite{majdabadi2020capsule} utilizes a two-layered capsule network with dynamic routing as the discriminator for its Multi-Scale Gradient capsule GAN, much like Sabour et al. Hsu et al.~\cite{8950449} based their work on the matrix capsules introduced by Hinton et al.~\cite{46653} to create two different approaches: one using capsules as its main component in the network and reconstructing HR images directly from it (CIRNN) and another using capsules for the channel attention mechanism (CARNN). Despite their solution performed better than other CNN methods, the baseline for their work were SR-CNN~\cite{7115171} and FSRCNN~\cite{10.1007/978-3-319-46475-6_25}, which were some of the first usage of convolutional neural networks for the SR problem. In this work, we employed more recent models for comparison.

\section{Proposed Method}
\label{sec:method}

\begin{figure*}[!htb]
\centering
\includegraphics[width=16.0cm]{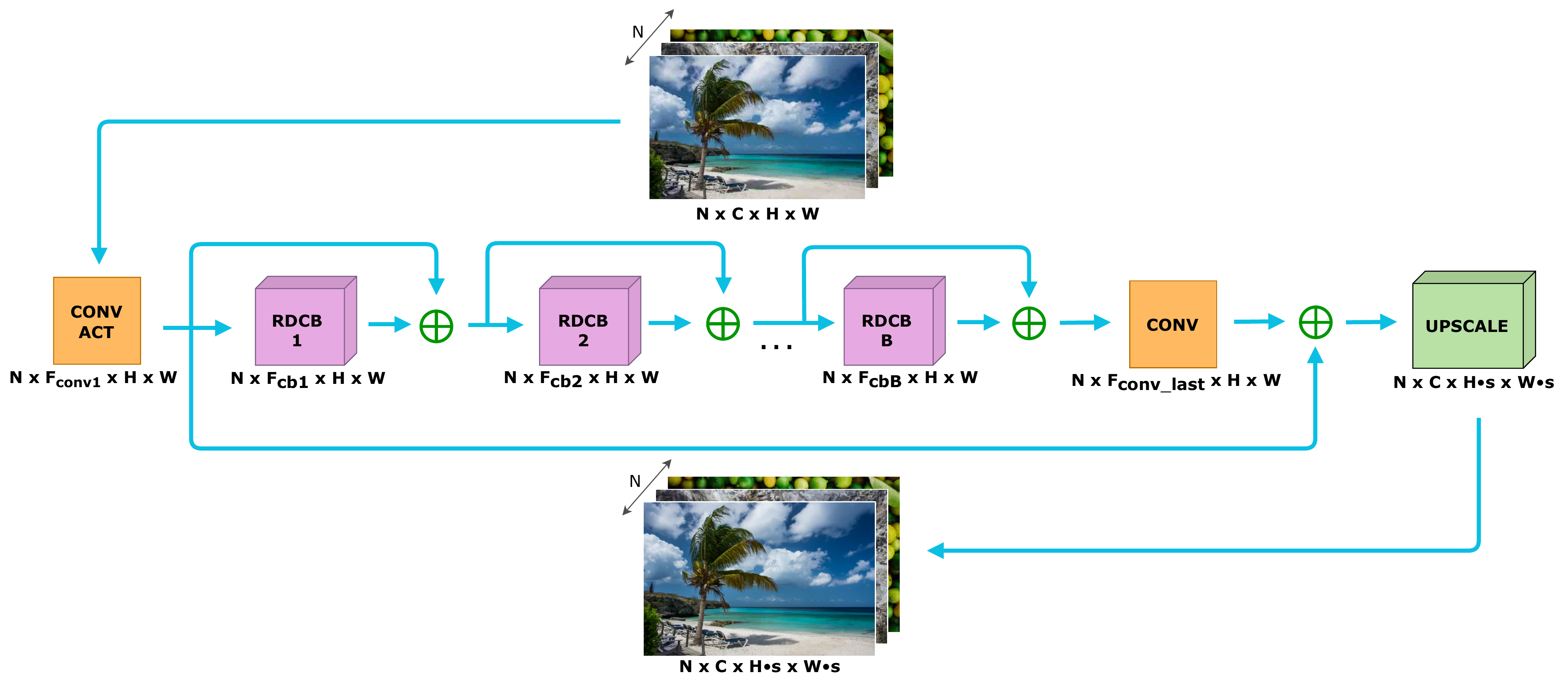}
\caption{Diagram of SRCaps model.}
\label{fig:model_diagram}
\end{figure*}

The proposed model, named Super-Resolution Capsules (SRCaps), is shown in Figure~\ref{fig:model_diagram}, which consists of four main parts: an initial convolutional layer, followed by $B$ sequentially connected residual dense capsule blocks, a new convolutional layer and, finally, a neural network to increase resolution. All the convolution-based layers use the weight normalization technique developed by Salimans and Kingma~\cite{NIPS2016_6114}, since this method accelerates the training convergence and has a lower computational cost if compared to batch normalization, without introducing dependencies between the examples of the batch, as shown in the work of Yu et al.~\cite{yu2018wide}.

The first convolutional layer, represented by the {\it CONV ACT} block in Figure~\ref{fig:model_diagram}, generates $F$ filters from convolutional kernels of size $k \times k$ with stride $st$ and padding $p$, followed by an activation function $act$. This layer is responsible for converting pixel intensities to local resource detector activations that are used as inputs to the next step in the capsule blocks.

\begin{figure*}[!htb]
\centering
\includegraphics[width=16.0cm]{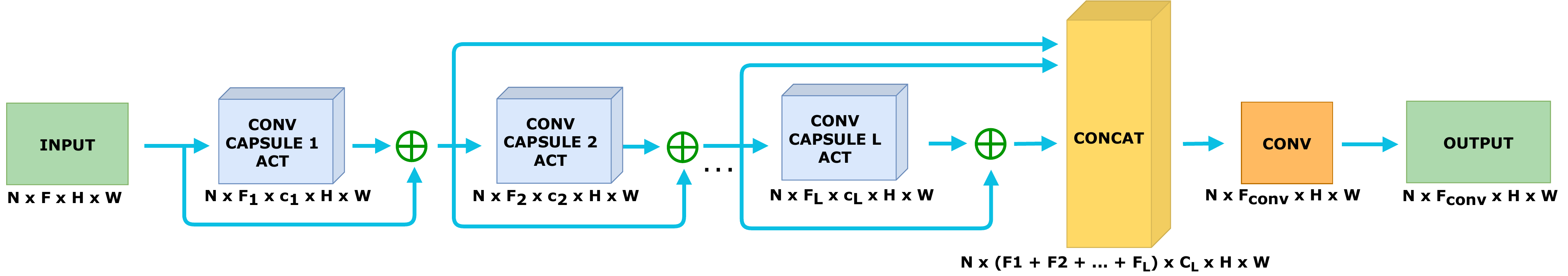}
\caption{RDCB diagram.}
\label{fig:capsblock_diagram}
\end{figure*}

The residual dense capsule blocks, hereinafter RDCBs, are composed of $L$ convolutional capsules layers followed by an activation function $act$, with residual connection to their inputs, sequentially connected. The outputs of these layers are concatenated, forming a dense connection, followed by a convolutional layer, as shown in Figure~\ref{fig:capsblock_diagram}. This convolutional layer, with kernels $1 \times 1$, stride 1 and padding 0, acts as a weighted sum between the various filters, allowing the network to learn which filters are more important, thus reducing dimensionality more efficiently, following its usage in Network in Network (NiN)~\cite{DBLP:journals/corr/LinCY13}, GoogLeNet~\cite{Szegedy_2015_CVPR}, VGGNet~\cite{Simonyan14c}, among other networks. The output of the RDCB is weighted by a residual scale constant $res\_scale$.

All capsules layers that are part of the same RDCB have the same parameters: the number of capsules per layer $c$, amount of filters $F$, kernel size $k$, stride $st$, and padding $p$.

At first, the routing algorithm used in the capsules was the same one proposed by~LaLonde and Bagci~\cite{lalonde2018capsules}, which not only drastically reduces the number of parameters and memory required in the original capsule definition but also allows the processing of larger images. This algorithm differs from the routing-by-agreement implementation by Sabour et al.~\cite{NIPS2017_6975} in two ways: capsules from the previous layer are only routed to capsules in the next layer within a specific spatial window, unlike the original algorithm that directs the output of all previous layer capsules to all capsules in the next layer, varying only the routing weight; and the transformation matrices are shared among all capsules of the same type.

In a later step, we decided to replace the initial routing algorithm for the no-routing introduced by Gu et. al.~\cite{Gu_2020_CVPR}. The authors argue that the routing procedure contributes neither to the generalization ability nor to the affine robustness of the CapsNets, concluding that different ways to approximate the coupling coefficients do not make a significant difference since they will be learned implicitly. In the no-routing approach, the iterative routing procedure is removed by setting all coupling coefficient as a constant $\frac{1}{M}$, where $M$ is the number of capsules in the next layer. We also tried different values for the squashing constant $sq$ used in the squashing function, as done by Huang and Zhou~\cite{huang2020capsnet}.

The RDCBs are sequentially connected, each having a residual connection with the block input, followed by a new convolutional layer, and the output of that layer has a residual connection with the output of the first convolutional layer. The existing residual connections, identified by the symbol $\varoplus$ in the model, were used for the following purposes: they avoid the problem of vanishing gradients (it becomes zero) by introducing shorter paths, which can take the gradient over the entire length of very deep networks, as demonstrated by~Veit et al.~\cite{NIPS2016_6556}; and the use of residual connections seems to greatly improve training speed, as observed by~Szegedy et al.~\cite{DBLP:journals/corr/SzegedyIV16}.

\begin{figure*}[!htb]
\centering
\includegraphics[width=14.0cm]{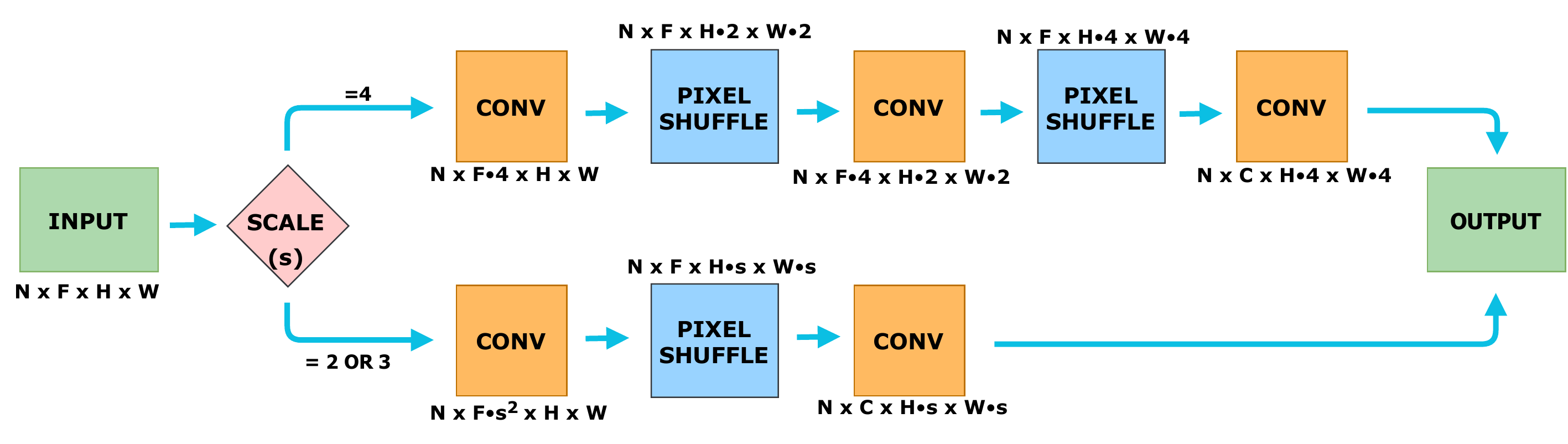}
\caption{UPNet diagram.}
\label{fig:upnet_diagram}
\end{figure*}

At the end of the model, there is a network used to perform resolution increase, called up-sampling network (UPNet) and shown in Figure~\ref{fig:upnet_diagram}. Following the works of Lim et al.~\cite{Lim_2017_CVPR_Workshops}, Zhang et al.~\cite{DBLP:journals/corr/abs-1802-08797}, and several participants of the NTIRE 2017~\cite{8014883} and NTIRE 2018~\cite{Timofte_2018_CVPR_Workshops} competitions, we chose for this purpose a network composed of subpixel convolutions, firstly introduced by Shi et al.~\cite{Shi_2016_CVPR}.

Using the UPNet built-in to the model allows the network to implicitly learn the process required to generate the larger version by adding the LR space feature maps and creating the SR image in a single step, saving memory and processing. This method was preferred over deconvolution since it naturally avoids checkerboard artifacts, which with deconvolution must be done using a kernel size that is divisible by stride to avoid the overlapping problem as demonstrated by Odena et al.~\cite{odena2016deconvolution}, and because it has a considerably lower computational cost, becoming $\log_{2}r^{2}$ times faster during training~\cite{Shi_2016_CVPR}.

\subsection{Loss Functions}
\label{subsec:funcoes_de_perda}

During training, we evaluated loss functions commonly used in solutions for super-resolution problems and also recently developed techniques, as well as functions for related problems. The most commonly employed function in the literature, and consequently in this work, is L1. It is widely used for being a simpler function and achieving satisfactory results. Zhao et al.~\cite{7797130} showed that a network trained with L1 has achieved superior results compared to the same network trained with L2, which was also observed by Lim et al.~\cite{Lim_2017_CVPR_Workshops}.

Still in the work of Zhao et al., the idea of using indices based on the Structural Similarity Index (SSIM)~\cite{1284395} is introduced for training neural networks. As previously noted by Dong et al.~\cite{10.1007/978-3-319-10593-2_13}, if a metric based on visual perception is used during training, the network can adapt to it. The SSIM is calculated as:
\begin{alignat}{2}
\label{eq:ssim}
\text{SSIM}(p) & =\frac{2 \mu_{x} \mu_{y}+C_{1}}{\mu_{x}^{2}+\mu_{y}^{2}+C_{1}} \cdot \frac{2 \sigma_{x y}+C_{2}}{\sigma_{x}^{2}+\sigma_{y}^{2}+C_{2}} \\
\label{eq:ssim2}
 & = l(p) \cdot cs(p)
\end{alignat}
\noindent where $\mu_{x}$ and $\mu_{y}$ are the average pixel values in the SR and HR patches, respectively, $\sigma_{x}$ and $\sigma_{y}$ are the standard deviations of the same patches, $\sigma_{x y}$ is the covariance between them, and $C_{1}$ and $C_{2}$ are constants added to avoid instabilities when the values of $\mu_{x}^{2}+\mu_{y}^{2}$ and $\sigma_{x}^{2}+\sigma_{y}^{2}$ are very close to $0$. The $l(p)$ part of the equation calculates the comparison between clipping luminances, while the comparison between their contrasts and structures is calculated by $cs(p)$. Since the highest possible value for SSIM is 1, and because training a neural network usually aims to minimize the loss function, we can define the $\mathcal{L}^{\text{SSIM}}$ function as:
\begin{equation}
\label{eq:ssim_loss}
\mathcal{L}^{\text{SSIM}}(P)=1-\text{SSIM}(\tilde{p})
\end{equation}
\noindent in which $\tilde{p}$ is the central pixel of patch $p$.

The best performance in the work of Zhao et al.~\cite{7797130} was obtained by combining L1 and the Multi-Scale Structural Similarity Index (MS-SSIM) shown in Equation~\ref{eq:ms-ssim_loss} weighing 0.16 and 0.84, respectively. The authors argue that MS-SSIM preserves contrast in high-frequency regions, while the L1 preserves color and brightness regardless of the local structure. The MS-SSIM value is obtained by combining measurements at different scales using the Equation:
\begin{equation}
\label{eq:ms-ssim}
\text{MS-SSIM}(p)=l_{M}^{\alpha}(p) \cdot \prod_{j=1}^{M} cs_{j}^{\beta_{j}}(p)
\end{equation}
\noindent where scales are used ranging from $1$ (original image) to $M$ (largest scale used), reducing the image by a factor of 2 every iteration; $l_{M}$ and $cs_{j}$ are the same terms as defined in Equation~\ref{eq:ssim2} at $M$ and $j$ scales, respectively, while the $\alpha$ and $\beta_{j}$ exponents are used to adjust the relative significance of different components. It is worth noting that the luminance comparison ($l_{M}$) is calculated only at $M$ scale, while contrast and structure comparisons ($cs_{j}$) at each scale. As with SSIM, the largest possible value for MS-SSIM is 1, so we can use it in a loss function in the form:
\begin{equation}
\label{eq:ms-ssim_loss}
\mathcal{L}^{\text{MS-SSIM}}(P)=1-\text{MS-SSIM}(\tilde{p})
\end{equation}

We also explored the combination of functions employing several different layers of the network, as in the work of Xu et al.~\cite{8014881}, in which the weighted sum between the calculation of the L1 function after two, three and four residual blocks are used, with weights of 0.5, 0.5 and 1, respectively. After each residual block that is employed in the loss calculation, a network based on subpixel convolutions is added to perform upscaling. Another possibility investigated was the use of edge maps in the error function, since the L1 error may smooth the edges. As in the work of Pandey et al.~\cite{DBLP:journals/corr/abs-1809-00961}, we evaluated a combination of the L1 using the SR and HR images and the L1 between its edge maps. However, although their work uses the Canny operator~\cite{4767851} to generate the edge map, our work investigated the usage of the Sobel operator~\cite{Sobel1968}.

Looking for alternatives in related problems, other investigated functions were the three-component weighted PSNR (3-PSNR) and the three-component weighted SSIM (3-SSIM), described in the work of Li and Bovik~\cite{doi:10.1117/1.3267087} and used to measure the quality of images and videos. This approach consists of breaking up an image into three parts: edges, textures, and more homogeneous regions. To do this, the Sobel operator is applied to the luminance channel of the image and, from the highest calculated value and some pre-established values, the thresholds that delimit each region are calculated.

The value of each of the metrics is calculated by applying different weights for each region: Li and Bovik~\cite{doi:10.1117/1.3267087} showed that the weights that achieved the best results were 0.7, 0.15 and 0.15 for the 3-PSNR, and of 1, 0 and 0 for 3-SSIM, considering edges, textures, and homogeneous regions, respectively. These values are consistent with the observation that perturbations at the edges of an object are perceptually more significant than in other areas.

Based on recent solutions available in the literature, Barron~\cite{DBLP:journals/corr/Barron17} presented a loss function that is a superset of Cauchy/Lorentzian, Geman-McClure, Welsch/Leclerc, generalized Charbonnier, Charbonnier/pseudo-Huber/L1-L2, and L2. This function has two hyperparameters: robustness ($\alpha$) and scale ($c$), with the variation of which is possible to reach all previous functions as specific cases. The general loss function is calculated as follows:
\begin{equation*}
\label{eq:barron}
\mathcal{L}^{\text{general}}(x, \alpha, c) \!=\! \left\{\begin{array}{ll}{\frac{1}{2}(x / c)^{2}} \!\!&\!\! {\mathrm { if }\ \alpha=2} \\
        {\log \left(\frac{1}{2}(x / c)^{2}+1\right)} \!\!&\!\! {\mathrm { if }\ \alpha=0} \\
        {1-\exp \left(-\frac{1}{2}(x / c)^{2}\right)} \!\!&\!\! {\mathrm { if }\  \alpha=-\infty} \\
        {\frac{|2-\alpha|}{\alpha}\left(\left(\frac{(x / c)^{2}}{|2-\alpha|}+1\right)^{(\alpha / 2)}-1\right)} \!\!&\!\! { \mathrm{ otherwise }}\end{array}\right.
\end{equation*}
\noindent where $x$ is the difference between the HR and SR pixel values. Barron also showed that it is possible to modify its function so that the network learns optimal values for the $\alpha$ and $c$ parameters, thus providing an appropriate exploration by the network of different error functions. Due to its unique features, we decided to use the adaptive loss function for the SRCaps training.


\subsection{Metrics}
\label{sec:metricas}

For the validation process of the results obtained, we employed metrics commonly used in the literature. More specifically, Peak Signal-to-Noise Ratio (PSNR)~\cite{5596999}, Structural Similarity Index (SSIM)~\cite{1284395} and Multi-Scale Structural Similarity Index (MS-SSIM)~\cite{1292216}.

The PSNR is a widely used metric in the quantitative evaluation of the quality of restoration, compression, and image construction because it is simple to calculate and easy to incorporate into the optimization process. Its calculation involves the mean of the square of the differences between the reference and generated pixels intensities, which is the same as the mean squared error (MSE), but applied only in the luminance channel. The MSE and PSNR equations are given as:
\begin{equation}
\label{eq:mse}
\mse = \frac{1}{HW} \sum_{i=0}^{H-1} \sum_{j=0}^{W-1}[I_{HR}(i, j)-I_{SR}(i, j)]^{2}
\end{equation}
\begin{equation}
\label{eq:psnr}
\psnr = 10 \cdot \log _{10}\left(\frac{255^{2}}{\mse}\right)
\end{equation}

The SSIM and MS-SSIM metrics, shown in Equations~\ref{eq:ssim} and~\ref{eq:ms-ssim}, respectively, are widely used because they are based on the assumption that the human visual system is highly adaptive for extracting structural information from the scene. Thus, these metrics are also employed because they provide a good approximation to the quality of the perceived image. The generated model was compared to models currently used as baseline, such as the winner of the NTIRE 2017 Super-Resolution Challenge, EDSR~\cite{Lim_2017_CVPR_Workshops}, as well as models that currently configure the state of the art in super-resolution, such as RDN~\cite{DBLP:journals/corr/abs-1802-08797}, RCAN~\cite{zhang2018rcan}, and WDSR~\cite{yu2018wide}, being the latter the winner of the NTIRE 2018 Challenge on Single Image Super-Resolution in all three realistic categories.

\begin{figure*}[!htb]
\centering
\begin{subfigure}{0.2\textwidth}
    \includegraphics[width=\linewidth]{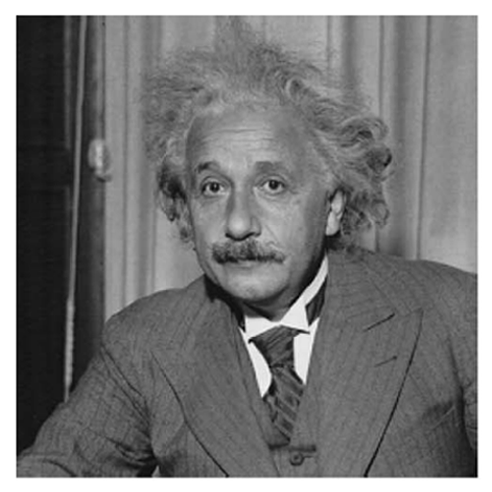}
    \caption{}
\end{subfigure}
\hspace*{0.3cm}
\begin{subfigure}{0.2\textwidth}
    \includegraphics[width=\linewidth]{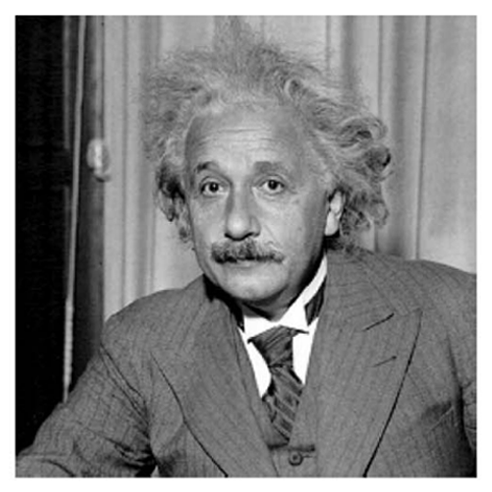}
    \caption{}
\end{subfigure}
\hspace*{0.3cm}
\begin{subfigure}{0.2\textwidth}
    \includegraphics[width=\linewidth]{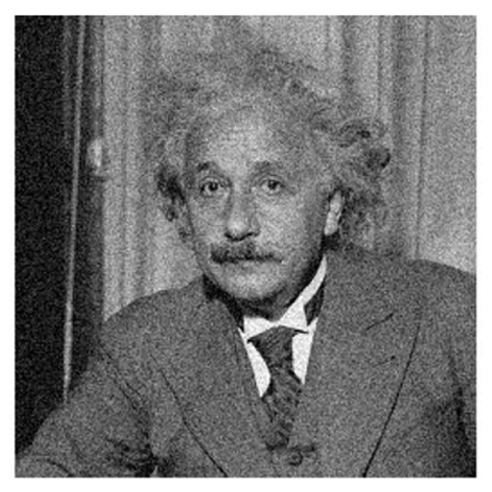}
    \caption{}
\end{subfigure}
\hspace*{0.3cm}
\begin{subfigure}{0.2\textwidth}
    \includegraphics[width=\linewidth]{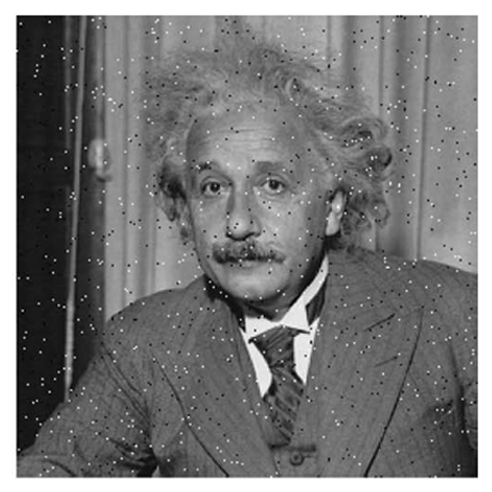}
    \caption{}
\end{subfigure}
\begin{subfigure}{0.2\textwidth}
    \includegraphics[width=\linewidth]{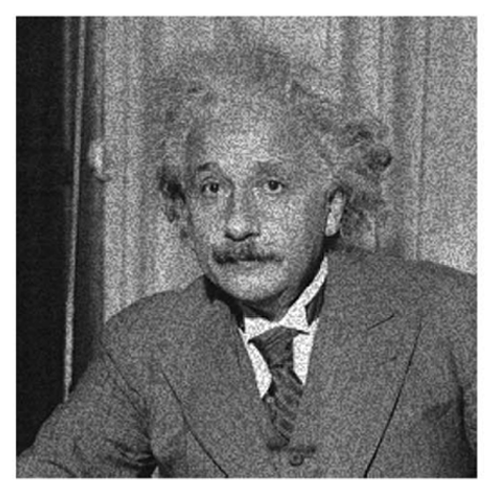}
    \caption{}
\end{subfigure}
\hspace*{0.3cm}
\begin{subfigure}{0.2\textwidth}
    \includegraphics[width=\linewidth]{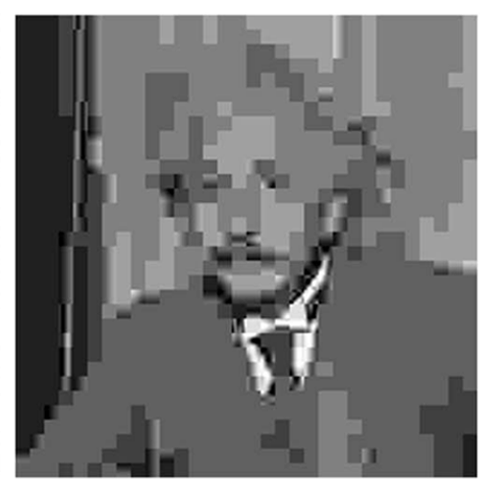}
    \caption{}
\end{subfigure}
\hspace*{0.3cm}
\begin{subfigure}{0.2\textwidth}
    \includegraphics[width=\linewidth]{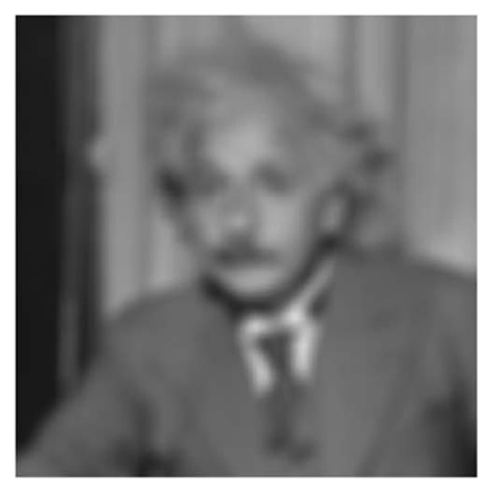}
    \caption{}
\end{subfigure}
\caption{Image produced by Li and Bovik~\cite{doi:10.1117/1.3267087} demonstrating images with same PSNR: (a) reference image, (b) contrast enhanced image, (c) Gaussian noise contaminated image, (d) salt-pepper noise contaminated image, (e) speckle noise contaminated, (f) JPEG compressed image, (g) blurred image.}
\label{fig:same_psnr}
\end{figure*}

\begin{figure*}[!htb]
\centering
\begin{subfigure}{0.23\textwidth}
    \includegraphics[width=\linewidth]{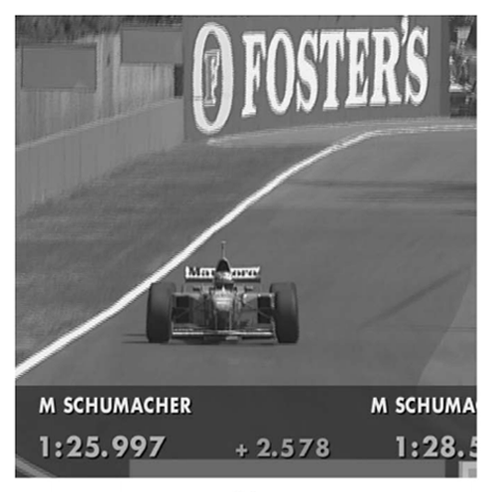}
    \caption{}
\end{subfigure}
\hspace*{1.0cm}
\begin{subfigure}{0.23\textwidth}
    \includegraphics[width=\linewidth]{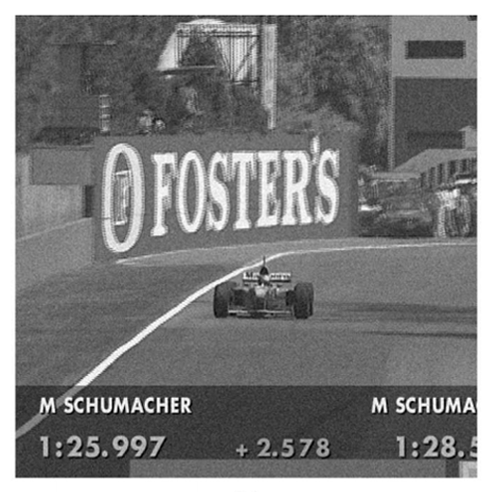}
    \caption{}
\end{subfigure}
\hspace*{1.0cm}
\begin{subfigure}{0.23\textwidth}
    \includegraphics[width=\linewidth]{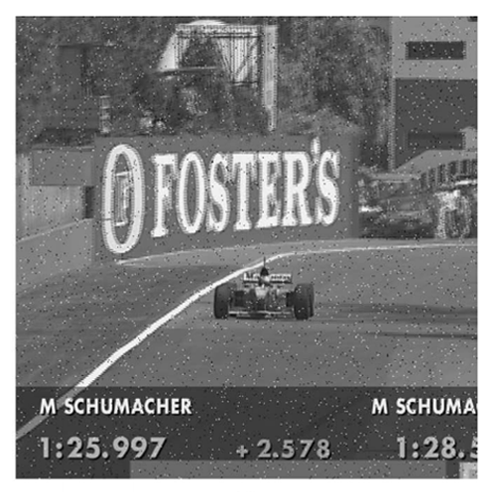}
    \caption{}
\end{subfigure}
\\
\begin{subfigure}{0.23\textwidth}
    \includegraphics[width=\linewidth]{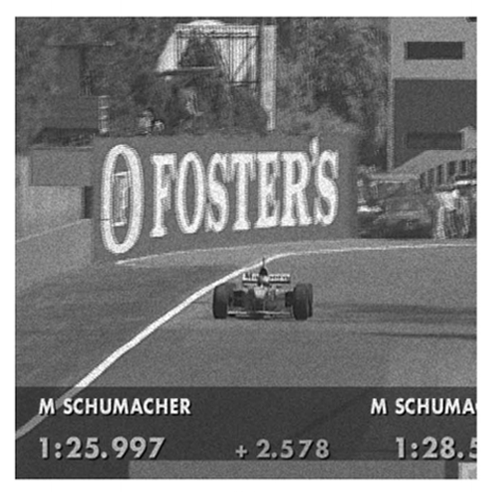}
    \caption{}
\end{subfigure}
\hspace*{1.0cm}
\begin{subfigure}{0.23\textwidth}
    \includegraphics[width=\linewidth]{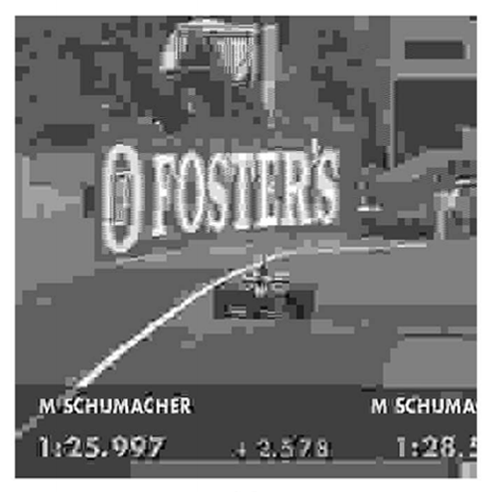}
    \caption{}
\end{subfigure}
\hspace*{1.0cm}
\begin{subfigure}{0.23\textwidth}
    \includegraphics[width=\linewidth]{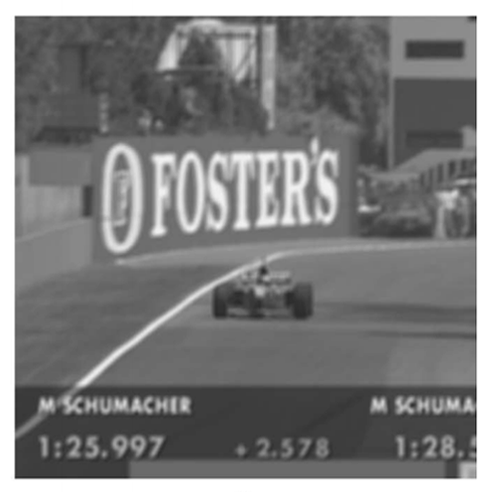}
    \caption{}
\end{subfigure}
\caption{Image produced by Li and Bovik~\cite{doi:10.1117/1.3267087} demonstrating images with same SSIM: (a) reference image, (b) Gaussian noise contaminated image, (c) salt-pepper noise contaminated image, (d) speckle noise contaminated, (e) JPEG compressed image, (f) blurred image.}
\label{fig:same_ssim}
\end{figure*}

It is worth noticing that, even though PSNR and SSIM are metrics widely used in the literature and their results indicate similarity between images, they cannot be taken as absolute truth when evaluated in the face of human perception. Examples of this fact have been demonstrated by Li and Bovik~\cite{doi:10.1117/1.3267087} and reproduced in this work. As exemplified in Figure~\ref{fig:same_psnr}, the metric fails to describe the similarity between the images, with all getting the same PSNR compared to the reference image. This fact had already been observed by other authors, as in the work of Wang et al.~\cite{1284395}, who developed a metric proposal closer to human perception. However, as shown in Figure~\ref{fig:same_ssim}, all the distorted versions of the image had the same SSIM in relation to the reference image, although some are visually inferior, indicating that it should not always be considered as truth for the perception assessment, which led this work also to consider the MS-SSIM metric and look for recent alternatives.

Algorithms that measure the differences between two images are usually developed assuming that the images are shown side by side or alternated with an empty image displayed in between for a short period before the next image is shown. In contrast, flipping (or alternating) between two similar images reveals their differences to an observer much more effectively than showing the images next to each other. Aiming to better approximate human evaluators' methods, Andersson et. al.~\cite{Andersson2020} developed a full-reference image difference algorithm, namely \reflectbox{F}LIP, which carefully evaluates differences inspired by models of the human vision.

\reflectbox{F}LIP is designed to have both low complexity and ease of use. It does not only evaluates differences in colors and edges, but also pays attention to discrepancies in isolated pixels with colors that greatly differ from their surroundings, while ignoring details that a human observer might not perceive. It outputs a new image indicating the magnitude of the perceived difference between two images at every pixel. The algorithm can also pool the per-pixel
differences down to a weighted histogram, or generate a single value, which is the approach we will use during our analysis. This value is zero when both images are the same, and it increases as the more noticeable are the differences.


\subsection{Datasets}
\label{sec:datasets}

In this work, we employed in the experiments a dataset that is widely used in the literature, following the most recent usages. Currently, the DIV2K training set is used in the training of neural networks for the super-resolution problem, while the validation set of DIV2K, B100, Set5, Set14, and Urban100 are used to validate the results. All datasets are composed of original versions of the images (HR) and their reduced versions by bicubic interpolation algorithm for $2\times$, $3\times$, and $4\times$. In this work we will focus on the $4\times$ scale factor.


\subsection{Computational Resources}

The proposed model, named Super-Resolution Capsules (SRCaps) and shown in Figure~\ref{fig:model_diagram}, was implemented using the PyTorch open source platform~\cite{paszke2017automatic} with the help of the PyTorch Lightning wrapper~\cite{falcon2019pytorch}. The implementation made available by Uchida\footnote{Available at \url{https://github.com/S-aiueo32/sr-pytorch-lightning}} was used as a foundation, and also metrics from the PyTorch Image Quality\footnote{Available at \url{https://github.com/photosynthesis-team/piq}} collection~\cite{piq} and the official implementation of the \reflectbox{F}LIP metric made available by the authors\footnote{Available at \url{https://research.nvidia.com/publication/2020-07_FLIP}}~\cite{Andersson2020}. For optimizers implementations we used the torch-optimizer package\footnote{Available at \url{https://github.com/jettify/pytorch-optimizer}} and for comparative visualization of the metrics and generation of graphs the Tensorboard~\cite{tensorflow2015-whitepaper} and Comet.ml\footnote{Available at \url{https://www.comet.ml/}} tools were used.

The entire training process was conducted on a wide variety of machines, with configurations ranging from an Intel\textsuperscript{\tiny\textregistered} Core\textsuperscript{\texttrademark} i7 3770K 3.50GHz processor, 32GB of RAM with Ubuntu 16.04 LTS operating system to dual Intel\textsuperscript{\tiny\textregistered} Xeon\textsuperscript{\texttrademark} Gold 6230 processors, 1TB of RAM with Ubuntu 20.04.1 LTS. These machines were used for non-exclusive processing (sometimes with processes that use another or even the same graphics card in the same machine) causing the division of its resources. They also had several models and amounts of installed graphics cards, ranging from models such as the NVIDIA GeForce GTX 1080p GPU, which has 11 GB of memory and compute capability of 6.1 to Quadro RTX 8000, which has 48 GB memory and compute capability of 7.5. The different graphics cards used and their configurations are summarized in Table~\ref{tab:summary_gpus}.

\begin{table*}[!htb]
\caption{Summary of differences from used NVIDIA GPUs.}
\label{tab:summary_gpus}
\setlength{\tabcolsep}{3.5mm}
\small
\centering
\begin{tabular}{lcc}
\toprule
GPU                 & Compute Capability    & Memory (GB) \\
\midrule
GeForce GTX TITAN X & 5.2                   & 12 \\
Tesla P100          & 6.0                   & 12 \\
TITAN X (Pascal)    & 6.1                   & 12 \\
GeForce GTX 1080 Ti & 6.1                   & 11 \\
TITAN Xp            & 7.0                   & 12 \\
TITAN V             & 7.0                   & 12 \\
Quadro RTX 8000     & 7.5                   & 48 \\
\midrule
\end{tabular}
\end{table*}

\section{Experimental Results}
\label{sec:results}

In this section, we present the experiments conducted to validate the proposed method and its results.


\subsection{Parameter Search}

To find the best set of parameters for our SRCaps model, we used the Ray~\cite{liaw2018tune} open source framework, more specifically Tune, which is a scalable hyperparameter tuning library built on top of Ray Core\footnote{Available at \url{https://docs.ray.io/en/master/tune/index.html}}. We employed the Async Successive Halving (ASHA) scheduler during the search, as it decides at each iteration which trials are likely to perform badly, and stops these trials, avoiding wasting resources on bad hyperparameter configurations. The list of possible values for each parameter is displayed in Table~\ref{tab:summary_params}. The different models were trained for at maximum 100 epochs in the DIV2K training dataset, and evaluated by their MS-SSIM performance on the first 10 images from the DIV2K validation dataset. These values were selected to allow fast experimentation of different sets of parameters, as it was observed during numerous training processes that the model tends to start stabilizing performance at around 100 epochs.

\begin{table*}[!htb]
\caption{Summary of different values tried for each parameter.}
\label{tab:summary_params}
\setlength{\tabcolsep}{3.5mm}
\small
\centering
\begin{tabular}{ll}
\toprule
Parameters                      & Values \\
\midrule
activation ($act$)              & Hardswish, LeakyReLU, Mish, PReLU, ReLU, TanhExp \\
batch size ($N$)                & 8, 16, 32 \\
capsules ($c$)                  & 2, 3, 4, 5, 6, 7, 8, 9, 10 \\
capsule blocks ($B$)            & 2, 3, 4, 5, 6, 7, 8, 9, 10 \\
capsule layers ($L$)            & 2, 3, 4, 5, 6, 7, 8, 9, 10 \\
filters ($F$)                   & 32, 64, 96, 128 \\
kernel size ($k$)               & 3, 5, 7, 9 \\
learning rate                   & 0.0001, 0.001, 0.01, 0.1 \\
losses                          & adaptive, adaptive + sobel, L1, L1 + sobel, 3-PSNR, 3-SSIM \\
optimizer                       & ADAM, RAdam, Ranger, SGD \\
patch size                      & 128, 192 \\
residual scale ($res\_scale$)   & 0.1, 0.25, 0.5, 0.75, 1 \\
squashing constant ($sc$)       & 0.25, 0.5, 0.75, 1 \\
\midrule
\end{tabular}
\end{table*}


\subsection{Experimental Setup}

During the training process of the different models used for comparison, the entries have $N = 16$ (batch size) pairs of image clippings from the dataset. For all models, during validation, the LR and HR images are used entirely one by one ($N = 1$). All models discussed here were evaluated for $4\times$ super-resolution for 2000 epochs, where each epoch involves only one iteration through the training dataset, and trained with two different loss functions: L1 and adaptive. Other functions have been evaluated, and their results will be briefly discussed throughout this section.

The final SRCaps model has convolutional kernels with $k = 3$ in its first layer, followed by 7 RDCBs ($B = 7$). The value used for the hyperparameters $L$ and $c$ are the same for all blocks: 3 and 4, respectively. In the last convolutional layer, we chose $k = 3$, as well as for the convolutional layers internal to the UPNet. Throughout the neural network, we used the values of $act = ReLU$, $k = 3$, $F = 128$, $st = 1$ and $p = \text{'same'}$. Setting the padding to the 'same' mode means using its value so that the input ($H$ and $W$) dimensions are preserved in the output, that is, $p = \frac{k-1}{2}$. The SRCaps network, although having a smaller number of layers, using only seven residual blocks in its composition, has a considerable number of parameters, as shown in Table~\ref{tab:parameters}. This is due to the vectorial nature of the capsule, which adds an extra dimension in its composition.

HR image slices ($patch\_size$) of size 128$\times$128 and its corresponding 32$\times$32 LR slice were used during their training. The updating of the weights of the nets is done by the Adam~\cite{DBLP:journals/corr/KingmaB14} optimizer, with $\beta_{1} = 0.9$, $\beta_{2} = 0.999$, and $\epsilon = 10^{-8}$, being the nets trained with an initial learning rate of $lr = 10^{-4}$ which decays to half of the current value every 500 epochs. A summary of the characteristics of the models is presented in Table~\ref{tab:parameters}.


\begin{table*}[!htb]
\caption{Characteristics of the models used for comparison.}
\label{tab:parameters}
\setlength{\tabcolsep}{3.5mm}
\small
\centering
\begin{tabular}{llllll}
\toprule
Characteristics             & \textbf{SRCaps}   & EDSR  & RCAN              & WDSR  & RDN \\
\midrule
Number of Parameters        & 15M               & 1.5M  & 12.6M             & 4.8M  & 22.3M \\
Number of Residual Blocks   & 7                 & 16    & $10 \times 16$    & 16    & 16 \\
Number of Layers per Block  & 4                 & 2     & 3                 & 3     & 8 \\
Dense Connections           & Yes               & No    & No                & No    & Yes \\
Uses mean RGB               & No                & Yes   & No                & Yes   & No \\
Sub-pixel Convolution       & Yes               & Yes   & Yes               & Yes   & Yes \\
Loss Function               & Adaptive          & L1    & L1                & L1    & L1 \\
\bottomrule
\end{tabular}
\end{table*}

When used, the adaptive~\cite{DBLP:journals/corr/Barron17} loss function was initialized with the default values from the official implementation\footnote{Available at \url{https://github.com/jonbarron/robust_loss_pytorch}.}. Employing these values is equivalent to starting the training with the Charbonnier/Pseudo-Huber function and letting the network learn from it what values for its parameters and, consequently, which function of the subset of the general function is more appropriate. The Charbonnier function is also known as Pseudo-Huber or L1-L2 because it behaves as L2 near the origin and as L1 elsewhere, and combines the sensitivity of L2 with the robustness of L1.

The EDSR model used is the base model defined by Lim et al.~\cite{Lim_2017_CVPR_Workshops}. We chose the simplest version of the model because it has a smaller number of parameters than the SRCaps model. It is composed of 16 residual blocks without residual scale application since only 64 feature maps (filters) are used per convolutional layer. All convolutions, including the ones internal to the upscale network, have a kernel size of 3$\times$3. During its execution, the input images are subtracted from the mean RGB values of the training images of the DIV2K dataset, which are 0.4488, 0.4371, and 0.4040. These values are in the scale 0 to 1 and are multiplied by the maximum pixel value, 255. The RDN~\cite{DBLP:journals/corr/abs-1802-08797} model, as well as the EDSR, generates 64 filters as the output of its convolutional layers, and uses $k = 3$ for all convolutional kernels, except for those used in the fusion layers of LFF and GFF, which have $k = 1$. In this model, 16 RDB blocks were used, with 8 convolutional layers each.

For the WDSR model, we used the original implementation by Yu et al.~\cite{yu2018wide}, made available from the authors in their Github\footnote{Available at \url{https://github.com/JiahuiYu/wdsr_ntire2018}}. The chosen version was \texttt{wdsr-b}, with 16 large residual blocks that generate 128 filters, but which internally generate convolutional layers with $6 \times$ more filters. This model, such as EDSR, also subtracts the mean RGB values from the DIV2K images.

The RCAN model we used is also the original implementation, developed by Zhang et al.~\cite{zhang2018rcan} and available in the same repository used as a baseline. It is composed of 10 residual groups (RG) that form the Residual in Residual (RiR) network structure, in which each RG is composed of 16 Residual Channel Attention Blocks (RCAB). It has $k = 3$ kernel sizes which generates $C = 64$ filters in all convolutional layers, except for those in the channel reduction and amplification mechanism, which have $k = 1$, and $\frac{C}{r}=4$ and $C = 64$ respectively, with reduction factor $r = 16$.


\subsection{Results}

\begin{figure*}[!htb]
\centering
\includegraphics[width=16.0cm]{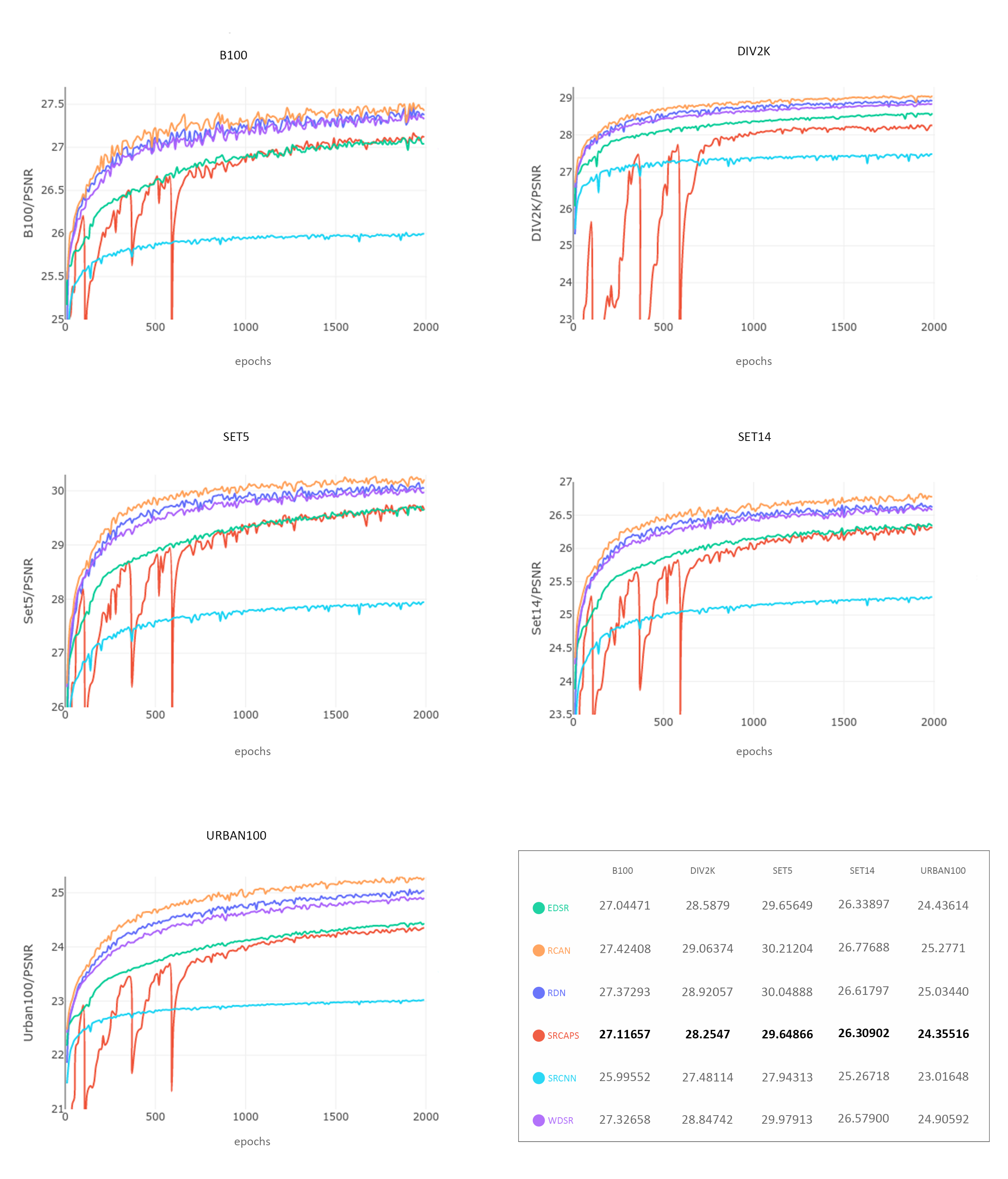}
\caption{Validation PSNR graphs.}
\label{fig:models_psnr}
\end{figure*}

\begin{figure*}[!htb]
\centering
\includegraphics[width=16.0cm]{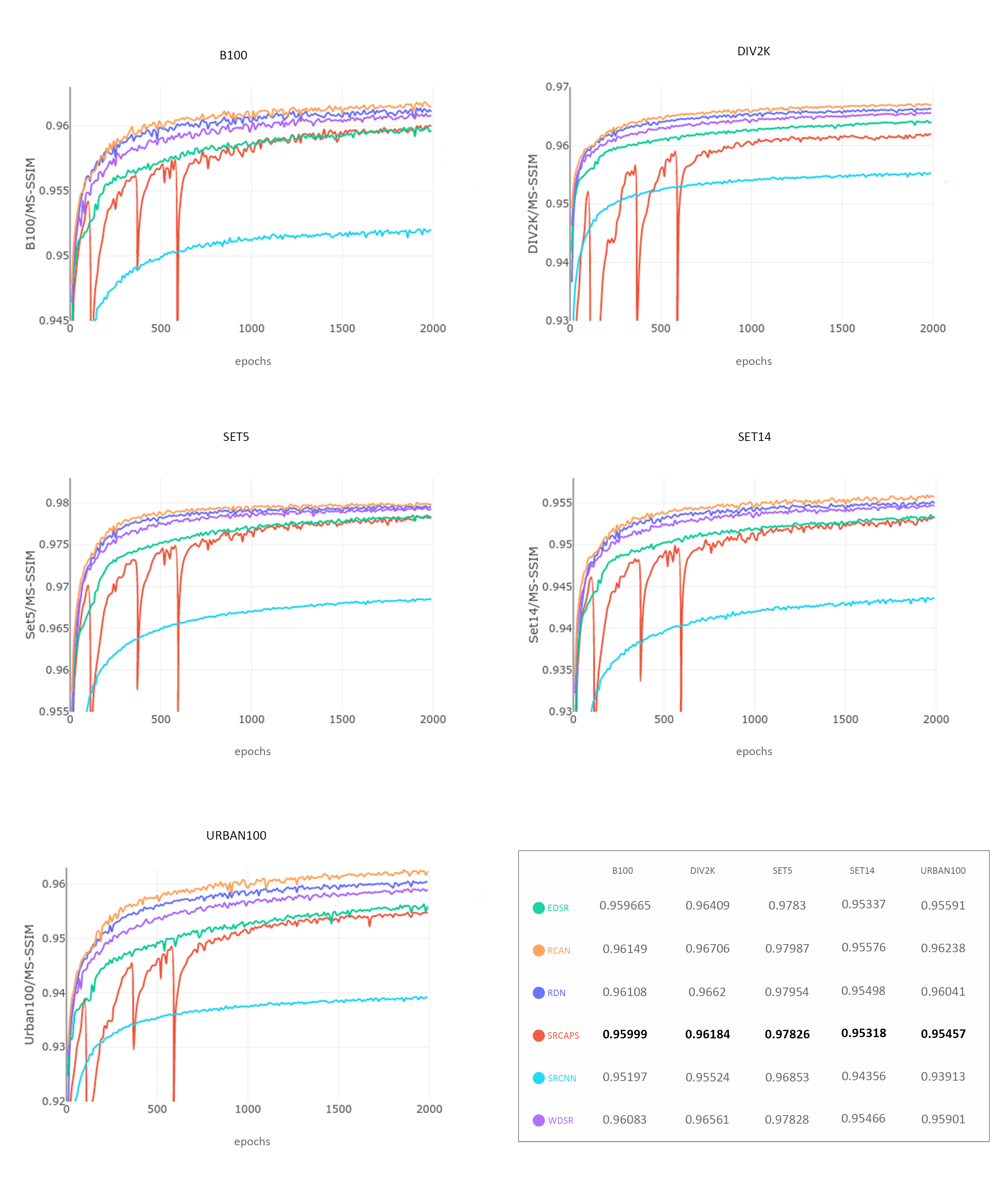}
\caption{Validation MS-SSIM graphs.}
\label{fig:models_msssim}
\end{figure*}

\begin{figure*}[!htb]
\centering
\includegraphics[width=16.0cm]{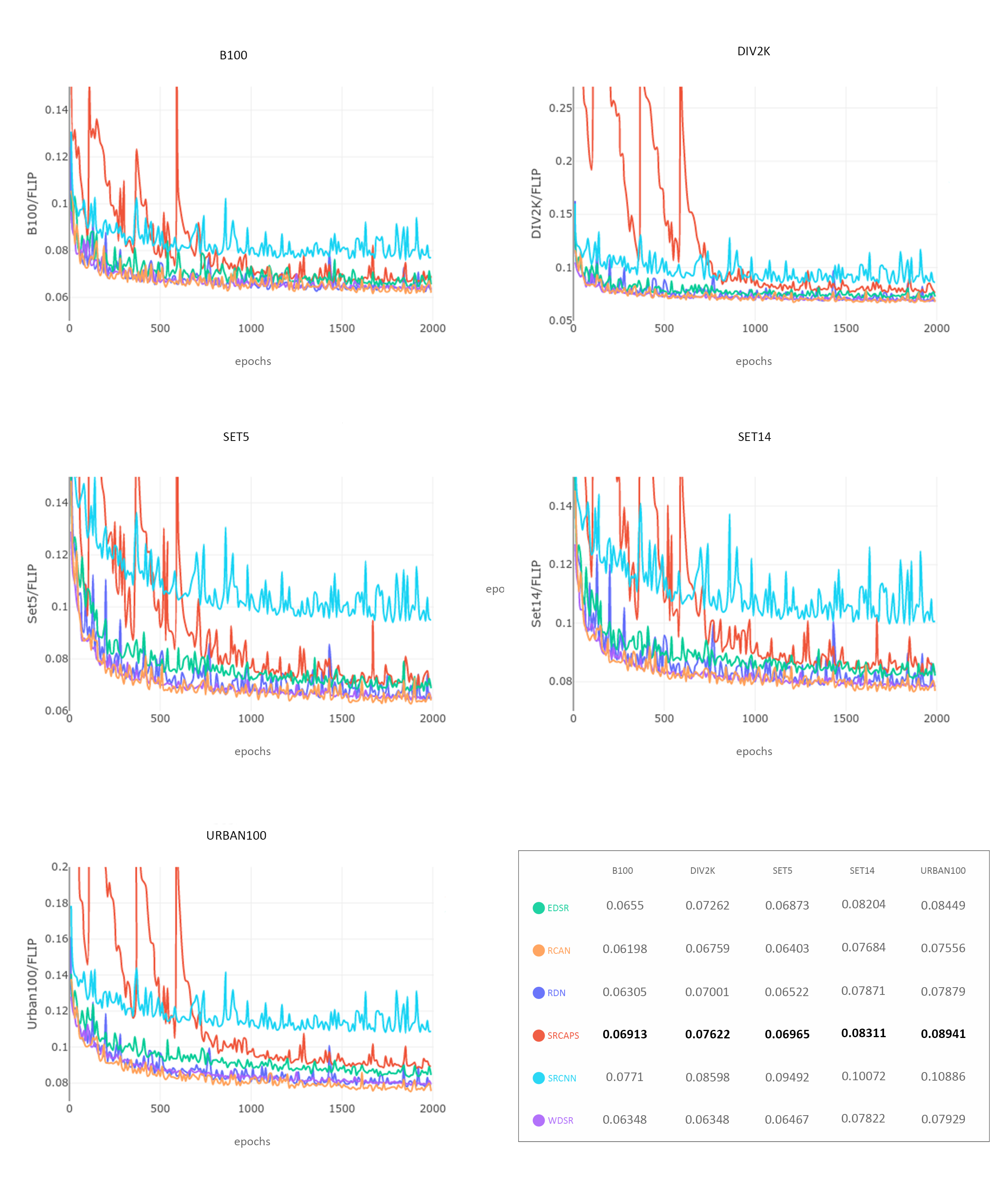}
\caption{Validation \protect\reflectbox{F}LIP graphs.}
\label{fig:models_flip}
\end{figure*}

It is possible to observe in Figures~\ref{fig:models_psnr}, \ref{fig:models_msssim}, and~\ref{fig:models_flip} that the evolution of the obtained result quality is similar in all metrics, forming graphs closer in shape. The developed model has a somewhat more chaotic behavior throughout the training due to the adaptive loss function, which learns along with the other network weights the best values for its parameters. The results obtained after 2000 training epochs for all models and datasets are shown in Table~\ref{tab:metrics}.

\begin{table*}[!htb]
\caption{Model metrics for each evaluated dataset. For the PSNR, SSIM, and MS-SSIM metrics the greater the better, while for the \protect\reflectbox{F}LIP metric the lesser the better. The best results for each metric and dataset is colored in blue.}
\label{tab:metrics}
\footnotesize
\centering
\begin{tabular}{|M{1.6cm}|M{1.5cm}|M{1.5cm}|M{1.2cm}|M{1.2cm}|M{1.2cm}|M{1.5cm}|M{1.2cm}|M{1.2cm}|N}
\hline
Datasets & Metrics & Bicubic & EDSR & RCAN & RDN & \textbf{SRCaps} & SRCNN & WDSR &\\[10pt] \hline \hline
B100 & \begin{tabular}[c]{@{}c@{}}PSNR\\ SSIM\\ MS-SSIM\\ \reflectbox{F}LIP\end{tabular} &
\begin{tabular}[c]{@{}c@{}}25.42985\\ 0.84128\\ 0.94109\\ 0.08295\end{tabular} &
\begin{tabular}[c]{@{}c@{}}27.04472\\ 0.78966\\ 0.95965\\ 0.06551\end{tabular} &
\begin{tabular}[c]{@{}c@{}}\textcolor{blue}{27.42409}\\ \textcolor{blue}{0.79804}\\ \textcolor{blue}{0.96149}\\ \textcolor{blue}{0.06198}\end{tabular} &
\begin{tabular}[c]{@{}c@{}}27.37293\\ 0.79646\\ 0.96108\\ 0.06306\end{tabular} &
\begin{tabular}[c]{@{}c@{}}27.11657\\ 0.78034\\ 0.95999\\ 0.06914\end{tabular} &
\begin{tabular}[c]{@{}c@{}}25.99553\\ 0.76109\\ 0.95198\\ 0.0771\end{tabular} &
\begin{tabular}[c]{@{}c@{}}27.32658\\ 0.79549\\ 0.96083\\ 0.06349\end{tabular} &
\\[30pt] \hline \hline
\begin{tabular}[c]{@{}c@{}}DIV2K\\ (validation)\end{tabular} &
\begin{tabular}[c]{@{}c@{}}PSNR\\ SSIM\\ MS-SSIM\\ \reflectbox{F}LIP\end{tabular} &
\begin{tabular}[c]{@{}c@{}}26.82531\\ 0.85062\\ 0.94505\\ 0.09288\end{tabular} &
\begin{tabular}[c]{@{}c@{}}28.5879\\ 0.80914\\ 0.9641\\ 0.07262\end{tabular} &
\begin{tabular}[c]{@{}c@{}}\textcolor{blue}{29.06375}\\ \textcolor{blue}{0.82104}\\ \textcolor{blue}{0.96706}\\ \textcolor{blue}{0.06759}\end{tabular} &
\begin{tabular}[c]{@{}c@{}}28.92058\\ 0.81771\\ 0.9662\\ 0.07001\end{tabular} &
\begin{tabular}[c]{@{}c@{}}28.25471\\ 0.79927\\ 0.96184\\ 0.07622\end{tabular} &
\begin{tabular}[c]{@{}c@{}}27.48115\\ 0.7788\\ 0.95525\\ 0.08599\end{tabular} &
\begin{tabular}[c]{@{}c@{}}28.84742\\ 0.8156\\ 0.96561\\ 0.06868\end{tabular} &
\\[30pt] \hline \hline
Set5 & \begin{tabular}[c]{@{}c@{}}PSNR\\ SSIM\\ MS-SSIM\\ \reflectbox{F}LIP\end{tabular} &
\begin{tabular}[c]{@{}c@{}}26.88569\\ 0.86689\\ 0.95716\\ 0.11142\end{tabular} &
\begin{tabular}[c]{@{}c@{}}29.6565\\ 0.8554\\ 0.97831\\ 0.06874\end{tabular} &
\begin{tabular}[c]{@{}c@{}}\textcolor{blue}{30.21205}\\ \textcolor{blue}{0.86464}\\ \textcolor{blue}{0.97988}\\ \textcolor{blue}{0.06404}\end{tabular} &
\begin{tabular}[c]{@{}c@{}}30.04888\\ 0.86269\\ 0.97954\\ 0.06523\end{tabular} &
\begin{tabular}[c]{@{}c@{}}29.64867\\ 0.85474\\ 0.97827\\ 0.06966\end{tabular} &
\begin{tabular}[c]{@{}c@{}}27.94313\\ 0.8119\\ 0.96853\\ 0.09493\end{tabular} &
\begin{tabular}[c]{@{}c@{}}29.97913\\ 0.86149\\ 0.97929\\ 0.06467\end{tabular} &
\\[30pt] \hline \hline
Set14 & \begin{tabular}[c]{@{}c@{}}PSNR\\ SSIM\\ MS-SSIM\\ \reflectbox{F}LIP\end{tabular} &
\begin{tabular}[c]{@{}c@{}}24.49373\\ 0.80909\\ 0.93117\\ 0.10949\end{tabular} &
\begin{tabular}[c]{@{}c@{}}26.33897\\ 0.74944\\ 0.95337\\ 0.08205\end{tabular} &
\begin{tabular}[c]{@{}c@{}}\textcolor{blue}{26.77688}\\ \textcolor{blue}{0.75981}\\ \textcolor{blue}{0.95576}\\ \textcolor{blue}{0.07684}\end{tabular} &
\begin{tabular}[c]{@{}c@{}}26.61797\\ 0.75649\\ 0.95498\\ 0.07872\end{tabular} &
\begin{tabular}[c]{@{}c@{}}26.30902\\ 0.74529\\ 0.95318\\ 0.08311\end{tabular} &
\begin{tabular}[c]{@{}c@{}}25.26719\\ 0.71373\\ 0.94356\\ 0.10072\end{tabular} &
\begin{tabular}[c]{@{}c@{}}26.579\\ 0.75594\\ 0.95467\\ 0.07823\end{tabular} &
\\[30pt] \hline \hline
Urban100 & \begin{tabular}[c]{@{}c@{}}PSNR\\ SSIM\\ MS-SSIM\\ \reflectbox{F}LIP\end{tabular} &
\begin{tabular}[c]{@{}c@{}}22.36726\\ 0.80426\\ 0.92388\\ 0.11929\end{tabular} &
\begin{tabular}[c]{@{}c@{}}24.43614\\ 0.77287\\ 0.95592\\ 0.0845\end{tabular} &
\begin{tabular}[c]{@{}c@{}}\textcolor{blue}{25.27712}\\ \textcolor{blue}{0.79915}\\ \textcolor{blue}{0.96239}\\ \textcolor{blue}{0.07557}\end{tabular} &
\begin{tabular}[c]{@{}c@{}}25.03441\\ 0.7915\\ 0.96041\\ 0.0788\end{tabular} &
\begin{tabular}[c]{@{}c@{}}24.35517\\ 0.76232\\ 0.95458\\ 0.08941\end{tabular} &
\begin{tabular}[c]{@{}c@{}}23.01649\\ 0.71431\\ 0.93913\\ 0.10886\end{tabular} &
\begin{tabular}[c]{@{}c@{}}24.90592\\ 0.78676\\ 0.95901\\ 0.07929\end{tabular} &
\\[30pt] \hline
\end{tabular}
\end{table*}

\begin{figure*}[!htb]
\centering
\includegraphics[width=16.0cm]{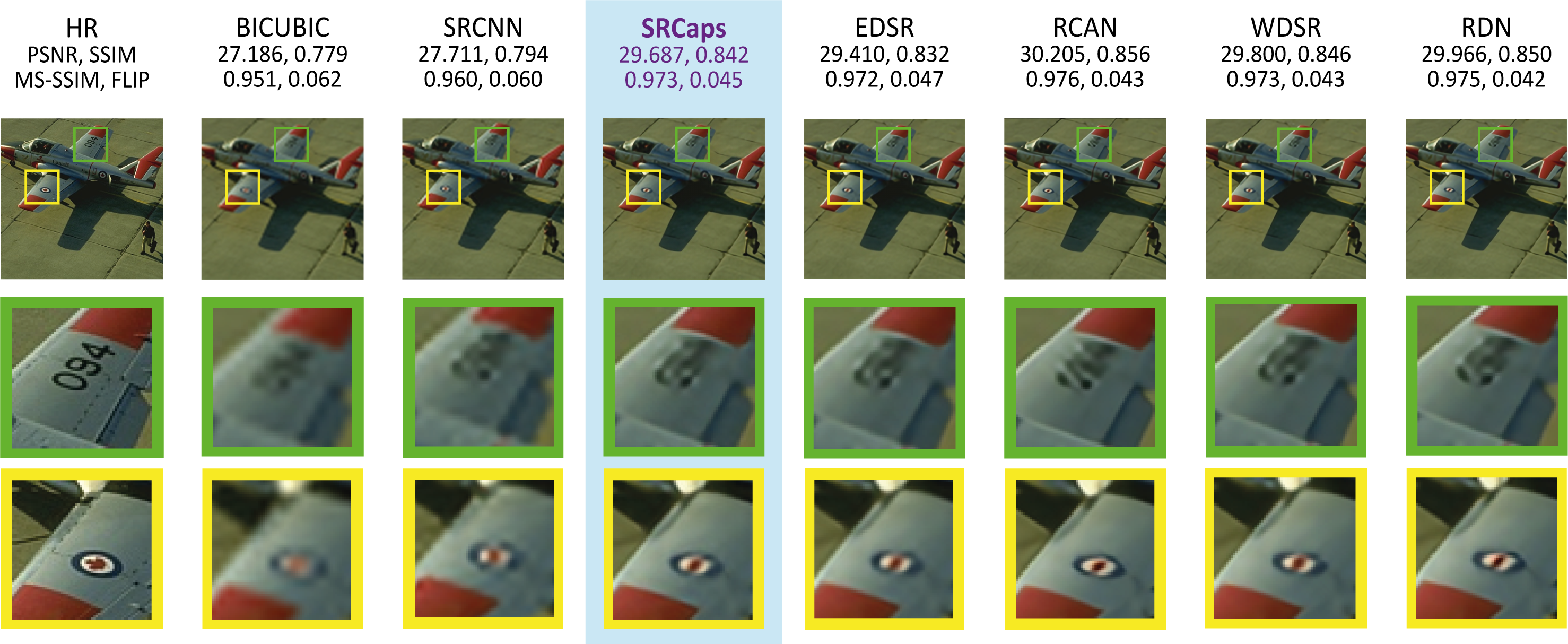}
\caption{Model results for ``37073'' image from B100 dataset.}
\label{fig:results_b100}
\end{figure*}

\begin{figure*}[!htb]
\centering
\includegraphics[width=16.0cm]{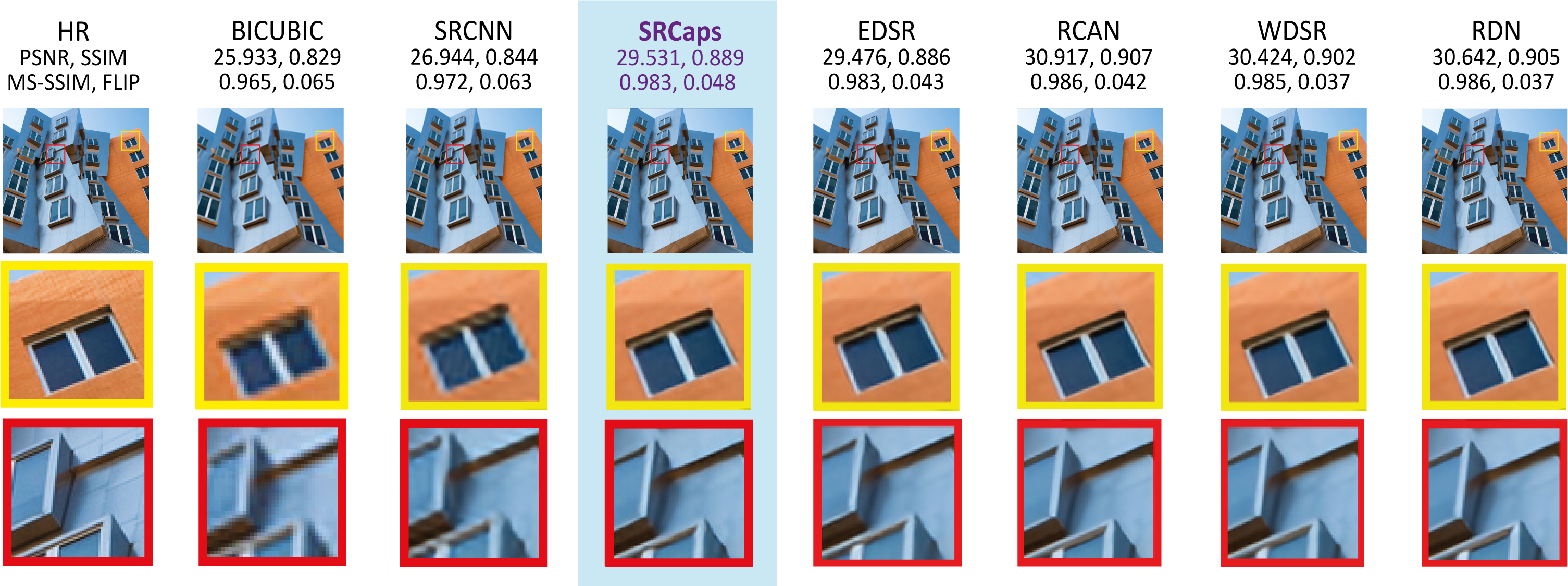}
\caption{Model results for ``img075'' image from Urban100 dataset.}
\label{fig:results_urban100}
\end{figure*}

It is remarkable the results obtained with the RCAN model, reaching the highest value in all metrics. It is also interesting to notice that the SRCaps model obtained results very close to the EDSR model, sometimes even surpassing it for some metrics, especially in the B100 dataset. Through the results shown in Figures~\ref{fig:results_b100} and~\ref{fig:results_urban100}, we can see the efficiency of the RCAN model, which manages to smooth the edges of the image so that it is visually pleasing.

\begin{figure*}[!htb]
\centering
\includegraphics[width=16.0cm]{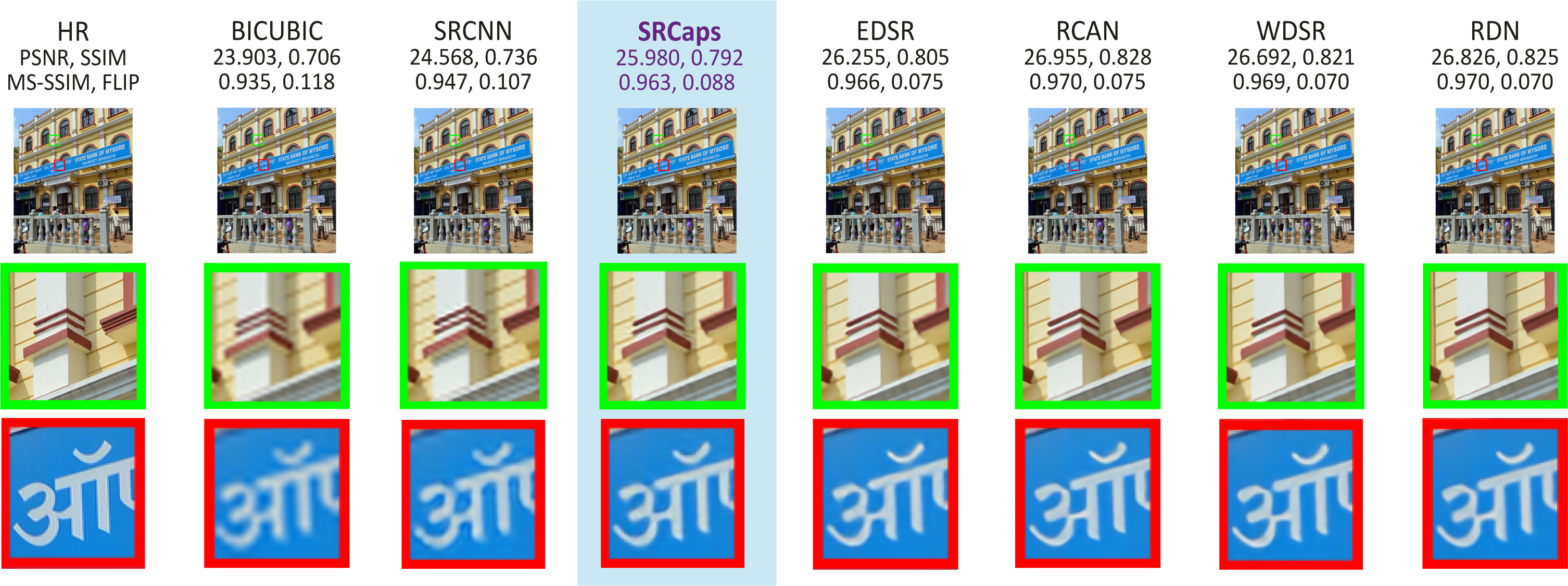}
\caption{Model results for ``0891'' image from DIV2K dataset.}
\label{fig:results_div2k_2}
\end{figure*}

\begin{figure*}[!htb]
\centering
\includegraphics[width=16.0cm]{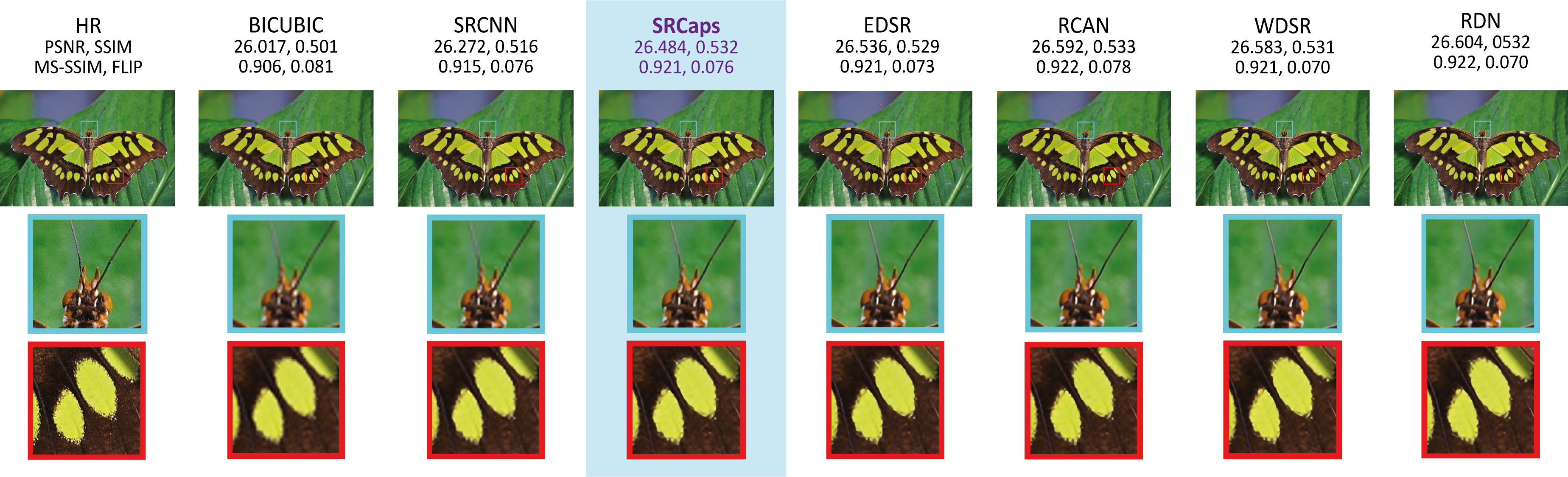}
\caption{Model results for ``0829'' image from DIV2K dataset.}
\label{fig:results_div2k_1}
\end{figure*}

By analyzing Figure~\ref{fig:results_div2k_2}, we can observe that the proposed model manages to re-create the connection between the different characters more precisely, while models with better metrics like RCAN and RDN tend to thin the connection, as they do for the leftmost part of the symbol on top. Other results can be visualized in Figure~\ref{fig:results_div2k_1}.

\begin{figure*}[!htb]
\centering
\includegraphics[width=16.0cm]{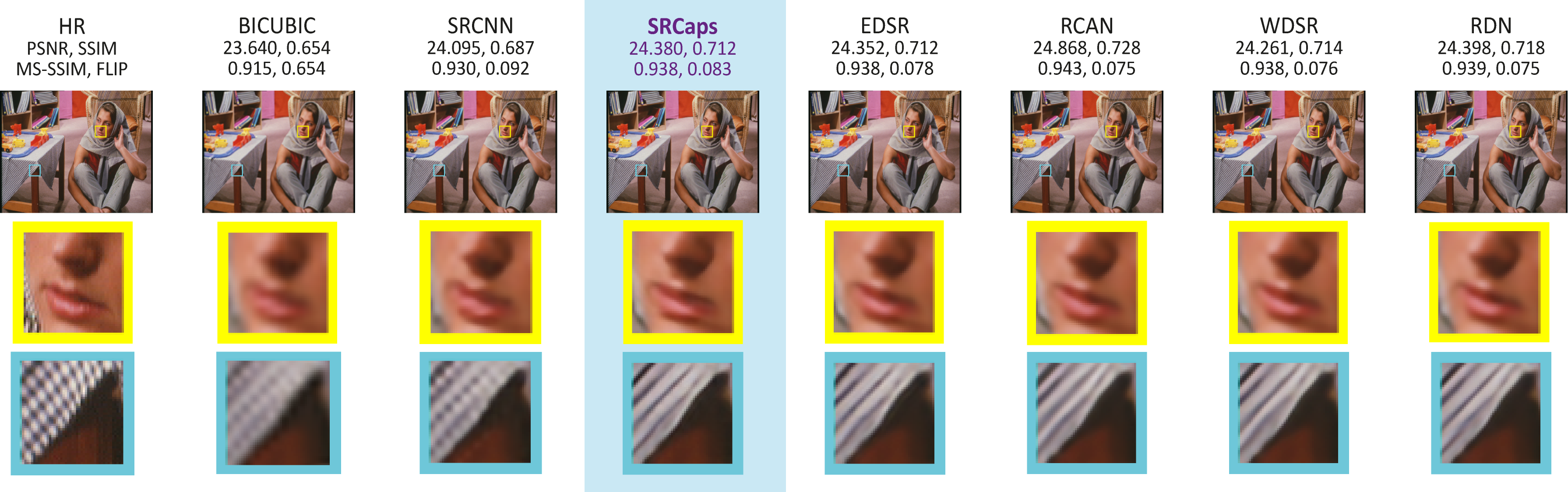}
\caption{Model results for ``barbara'' image from Set14 dataset.}
\label{fig:results_set14}
\end{figure*}

It is worth noticing that there are still room for improvement for all the analyzed models: in images such as Figure~\ref{fig:results_set14}, in which there is a checkerboard tablecloth, for softening the edges too much, the models turn the checkerboard pattern into parallel lines.

Despite being able to reconstruct with quality rounded edges, a deficiency of the SRCaps model is in the reconstruction of linear edges, usually diagonal. Aiming to solve this problem, several variations of the model were built, based on solutions previously developed in the literature, and that involved changes both in the architecture as a whole and in specific parts, as in the loss function, use of UPNet, change in the layers' inputs and even the network structure.

Inspired by the solution proposed by Wang et al.~\cite{DBLP:journals/corr/abs-1804-02900}, still during the elaboration of the presented model, experiments were carried out in which the results of each UPNet step that increases the image scale by $2\times$ is added to the result of an upscaling of the LR image by $2\times$ and $4\times$, respectively, performed by traditional interpolation methods. Also involving the upscaling was the idea of the model presented by Xu et al.~\cite{8014881}, where the output of several residual blocks is used in the calculation of the network loss. In their model, the outputs of the second, third and fourth blocks generate SR images, from which the L1 function is calculated in relation to the HR image, and its results are summed with weights 0.5, 0.5 and 1.0, respectively. These experiments returned lower results than the one presented.

Some loss function variations were evaluated, aiming to improve the generation of linear edges. The use of the 3-SSIM~\cite{doi:10.1117/1.3267087} as a loss function, both alone and in conjunction with the L1 and adaptive functions, was verified. Based on the work of Pandey et al.~\cite{DBLP:journals/corr/abs-1809-00961}, the L1 function was also used with the edge map of both SR and HR images, together with the same function calculated with the RGB images. The edge maps were generated from three different filters: Sobel operator, a filter created to identify diagonal edges and one to identify corners, but the results were unsatisfactory.

Changes in the network involving intermediate layers were also evaluated, however, without success. Among the tests performed is the use of the CoordConv concept, introduced by Liu et al.~\cite{DBLP:journals/corr/abs-1807-03247}, in which extra channels are added to the intermediate representations to allow the convolution access to their coordinates. This was done in the expectation that CoordConv would allow spatial information to be also encoded in the network, influencing more elongated edges. Another examined concept was the partial convolution based padding, created by Liu et al.~\cite{DBLP:journals/corr/abs-1811-11718}, in which the border padding that allows maintaining the proportion of the input image at the output of each module is treated as a hole filling problem, where the holes are the fill regions and the no hole areas are the original image.

As a preprocessing step, some alternatives involving both the creation of new layers and changes in the HR image were evaluated and discarded. In the addition of new layers, simpler solutions were evaluated, such as creating a layer with the edge maps of the input LR image through the Sobel operator, as well as more complex solutions involving the addition of three layers, each referring respectively to region of edges, textures and more homogeneous regions, as defined by Li and Bovik~\cite{doi:10.1117/1.3267087}. As for changes in the HR image, the use of the unsharp mask was evaluated before the loss function was calculated, to generate an image with more emphasis on the edges.

\section{Conclusions}
\label{sec:conclusions}

The purpose of this work was to evaluate the use of the capsule concept in the solution of single image super-resolution problems, as well as to verify new forms of training and validating the results of neural networks for this purpose. It was evidenced that, despite the inferior result, a trained network with a smaller number of layers obtained a relevant result, indicating that networks that use capsules can have applications in super-resolution. Hypotheses have been raised that the nonlinearity function applied together with the capsules may be a limiting factor, given the different nature of the problem as to its initial usage (super-resolution $\times$ classification).

Several loss function combinations were performed to improve the quality of the network training, as done in the work of Zhao et al.~\cite{7797130}. Functions widely used in super-resolution problems, such as L1, have been applied in conjunction with functions that take the human visual system into consideration, such as the SSIM~\cite{1284395} and MS-SSIM~\cite{1292216} functions. Further combinations using other functions described in the literature were also evaluated, such as 3-SSIM~\cite{doi:10.1117/1.3267087} and edge mapping functions~\cite{DBLP:journals/corr/abs-1809-00961}. The calculation of network loss from different layers was also evaluated, but with inexpressive results.

The fact that the adaptive function~\cite{DBLP:journals/corr/Barron17} is a superset of several others and that it is possible to make the network learn, along with the other weights, the optimal values for its two main parameters ($\alpha$ and $c$), allow the network to experiment which loss function best fits the problem. Thus, it is possible to train the network starting from a function similar to L1, while modifying it at each iteration to extract as much useful information as possible from the training data.

The current limitations of the most used metrics in the literature, based on previous studies~\cite{doi:10.1117/1.3267087}, were also emphasized in this work. Cases that demonstrate the inefficiency of the PSNR and SSIM metrics have been replicated, showing that visual evaluation of the results is still essential. Existing metrics in the literature have been suggested as an addition to those currently used, such as MS-SSIM~\cite{1292216} and \reflectbox{F}LIP~\cite{Andersson2020}, encouraging discussion of new metrics.

Throughout the development of this work, several points of possible improvements that could not be deeply evaluated were identified. Suggestions for future work permeate various parts of the created model. An initial idea would be to change the composition of the UPNet network, which is used in much the same way by several networks that have reached the state of the art. One can again verify the usage of the concept of reverse convolutions, or deconvolutions, as used by Dong et al.~\cite{10.1007/978-3-319-46475-6_25}, and also of deconvolutional capsules created by LaLonde and Bagci~\cite{lalonde2018capsules}, or use more recent methods from the literature. Kim and Lee~\cite{Kim_2018_CVPR_Workshops} recently proposed the enhanced upscaling module (EUM), which achieves better results through nonlinearities and residual connections.

Another suggestion would be to investigate new nonlinearity and routing functions for the capsules. It is assumed that the functions developed in the original work of Sabour et al.~\cite{NIPS2017_6975}, whose objectives are that the length of the output vector of a capsule indicates the probability that the entity represented by it is present in the current input, may not be the most appropriate for super-resolution problems, since the objective is not to identify class instances. One can also verify the use of other capsule models, such as the one developed by Hinton et al.~\cite{46653}, in which the capsules assume a matrix format rather than vectorial. The possible disadvantage of models in this format is the memory cost required by these capsules, which limits growth in both depth and width.

\section*{Acknowledgments}
\label{acknowledgment}

The authors would like to thank National Council for Scientific and Technological Development (CNPq \# 309330/2018-1) for its financial support and NVIDIA for the donation of a GPU as part of the GPU Grant Program.

\bibliographystyle{elsarticle-num}
\bibliography{IEEEabrv,paper}

\end{document}